\documentclass[aps,prb,showpacs,superscriptaddress
 ,twocolumn
]{revtex4}
\usepackage[dvips]{graphicx}
\usepackage{amsmath,amsfonts,amssymb}
\usepackage{multirow}
\usepackage{times}

\begin{document}
\author{B. A. Khoruzhenko}
 \affiliation{Queen Mary University of London, School of Mathematical Sciences, London E1 4NS, UK}
\author{D. V. Savin}
 \affiliation{Department of Mathematical Sciences, Brunel University, Uxbridge UB8 3PH, UK}
\author{H.-J. Sommers}
 \affiliation{Fachbereich Physik, Universit\"at Duisburg-Essen, 47048 Duisburg, Germany}
\title{Systematic approach to statistics of conductance and shot-noise in chaotic cavities}
\date{31 May 2009}

\begin{abstract}
Applying random matrix theory to quantum transport in chaotic cavities, we develop a novel approach to computation of the moments of the conductance and shot-noise (including their joint moments) of arbitrary order and at any number of open channels. The method is based on the Selberg integral theory combined with the theory of symmetric functions and is applicable equally well for systems with and without time-reversal symmetry. We also compute higher-order cumulants and perform their detailed analysis. In particular, we establish an explicit form of the leading asymptotic of the cumulants in the limit of the large channel numbers.  We derive further a general Pfaffian representation for the corresponding distribution functions. The Edgeworth expansion based on the first four cumulants is found to reproduce fairly accurately the distribution functions in the bulk even for a small number of channels. As the latter increases, the distributions become Gaussian-like in the bulk but are always characterized by a power-law dependence near their edges of support. Such asymptotics are determined exactly up to linear order in distances from the edges, including the corresponding constants.
\end{abstract}


\maketitle

\section{Introduction}

Quantum transport in mesoscopic systems has been a subject of an intense study during the last decade, see reviews \cite{Beenakker1997,Blanter2000}. Traditionally, the focus of interest has been on statistical properties of the conductance, $g$, and shot-noise, $p$. For noninteracting electrons, the Landauer-B\"uttiker scattering formalism \cite{Landauer} relates these quantities (expressed in their natural units) to the so-called transmission eigenvalues $T_i$ of the conductor:
\begin{equation}\label{g,p}
 g = \sum_{i=1}^{n} T_i\,, \qquad p = \sum_{i=1}^{n} T_i(1-T_i).
\end{equation}
Here, $n\equiv\mathrm{min}(N_1,N_2)$, where $N_{1,2}$ are the number of propagating modes (channels) in the two attached leads. The $T_i$ are the eigenvalues of the matrix $tt^{\dag}$, with $t$ being $N_1\times N_2$ matrix of transmission amplitudes from the entrance to exit channels. They are mutually correlated random numbers, $0 \leq T_i \leq 1$, whose distribution depends on the type of the conductor.

Below, we consider chaotic cavities (open quantum dots). In this case, random matrix theory (RMT) has proved to be successful in describing universal fluctuations in transport through such systems. \cite{Beenakker1997,Alhassid2000} Within this RMT approach, the joint probability density function (JPDF), $\mathcal{P}_\beta(T)$, of the transmission eigenvalues is induced by the random scattering matrix drawn from one of Dyson's circular ensembles \cite{Mehta2}, according to the global symmetries present in the system. The exact expression for this JPDF is known \cite{Beenakker1997,Baranger1994,Jalabert1994} to have the following simple form:
\begin{equation}\label{jpd}
  \mathcal{P}_\beta(T) = \mathcal{N}^{-1}_\beta |\Delta(T)|^{\beta}\prod_{i=1}^n
  T_i^{\alpha -1}\,,
\end{equation}
where $\alpha=\frac{\beta}{2}(|N_1-N_2|+1)$ and $\Delta(T)= \prod_{i<j}(T_i - T_j)$ denotes the Vandermonde determinant. The Dyson's symmetry index $\beta$ depends on the presence ($\beta=1$) or absence ($\beta=2$) of time-reversal symmetry or that of spin-flip symmetry ($\beta=4$) in the system, thus distinguishing between the three canonical RMT ensembles (orthogonal, unitary or symplectic, respectively). \cite{Mehta2}
The normalization constant $\mathcal{N}_\beta $ is given by
\begin{equation}\label{Nbeta}
  \mathcal{N}_\beta  = \prod_{j=0}^{n-1} \frac{\Gamma (1+ \frac{\beta}{2} (1+j)) \;
  \Gamma (\alpha + \frac{\beta}{2} j)\; \Gamma (1 + \frac{\beta}{2} j) }{
  \Gamma (1 + \frac{\beta}{2})\; \Gamma ( 1+ \alpha + \frac{\beta}{2} (n+j-1))}
\end{equation}
and assures that expression (\ref{jpd}) is a probability density. It is known generally for discrete positive $n$ and continuous $\alpha$ and $\beta$ as the Selberg integral. \cite{Mehta2}

Presently, there is a substantial progress in the understanding of statistics of various transport observables in chaotic cavities that is due to the recent developments of new analytical methods in the theory. Among them, the Selberg integral theory plays a special and important role, see the recent review \cite{Forrester2008} for its current status. In the present context, it has been initially applied by two of us \cite{Savin2006} to find the average value of shot-noise and hence the Fano factor exactly. This approach has then been developed further to study full counting statistics of charge transfer \cite{Novaes2007} as well as to obtain exact explicit expressions for the shot-noise variance and for the skewness and kurtosis of the charge and conductance distributions \cite{Sommers2007a,Savin2008} (see also Refs. \cite{Brouwer1996,Araujo1998,Blanter2001i,Bulashenko2005,Bulgakov2006i,Gopar2006,Beri2007,Vivo2008a} for other RMT results on the relevant statistics in chaotic cavities). Since $\beta$ enters the Selberg integral as a continuous parameter, this method allows us to treat all the three ensembles on equal footing, thus giving a powerful alternative to diagrammatic \cite{Brouwer1996,Bulgakov2006i} or orthogonal polynomial \cite{Araujo1998,Vivo2008a} approaches, especially when the channel numbers are small.

A completely alternative treatment has been recently undertaken within the semiclassical approach  \cite{Richter2002,Braun2006} which represents quantum transport in terms of classical trajectories connecting the leads. By constructing  asymptotic semiclassical expansions for transport observables, this approach successfully accounts for both system-specific and universal (RMT) features, see Ref.  \cite{Mueller2007} for a review.

The case of $\beta{=}2$ (broken time-reversal symmetry) is known for several reasons to be the special one in RMT. For the problem in question, further progress in this case has been made very recently along the following two directions. Novaes \cite{Novaes2008} combined the Selberg integral with facts from the theory of symmetric functions to compute non-perturbatively moments of both the transmitted charge and conductance but not those of shot-noise. Alternatively, Osipov and Kanzieper \cite{Osipov2008} combined the theory of integrable systems with RMT, as given by (\ref{jpd}), bringing out an effective formalism for calculating the cumulants of the conductance and shot-noise. \cite{Osipov2009} However, the relevant consideration for the systems with preserved time-reversal symmetry, $\beta{=}1$, is still lacking.

The distribution functions of the conductance and shot-noise are also studied quite intensively on their own. However, no explicit expressions have been reported so far except for a few cases, namely, for the conductance distribution at $N_{1,2}=1,2$ \cite{Baranger1994,Jalabert1994,Garcia-Martin2001,Bulgakov2006i} and for the shot-noise distribution at $N_{1,2}=1$. \cite{Pedersen1998} Asymptotic analysis of the both distributions at $N_{1,2}\gg1$ has been performed very recently in Ref.~\cite{Vivo2008}, see also Ref.~\cite{Osipov2008}. To the best of our knowledge, no general results valid at arbitrary $N_{1,2}$ and $\beta\neq2$ are available thus far. Meanwhile, the conductance distribution with dephasing \cite{Brouwer1997ii} has been directly measured in open quantum dots \cite{Huibers1998b} and in microwave billiards.\cite{Hemmady2006b} The shot-noise power in chaotic cavities has been recently studied experimentally. \cite{Oberholzer2001} Counting electrons in quantum dots is also experimentally accessible. \cite{Gustavsson2009} All this provides an additional motivation for the present study.

In this work we explore further the direction along the lines of Novaes's work \cite{Novaes2008} and develop a systematic approach for computing the moments of linear statistics in transmission eigenvalues for the systems with both preserved and broken time-reversal symmetry. This approach yields the moments of the conductance and shot-noise of arbitrary order, including their joint moments and cumulants. In the next section we present the detailed exposition of the method used, including the relevant facts from the theory of symmetric functions. This method is then applied in Sec. \ref{Moments} to derive expressions for the moments and cumulants of the conductance and shot-noise in a closed form. Sections \ref{Distributions} and \ref{Asymptotic} complement this study by investigating the corresponding distribution functions and their asymptotic behaviour. Our main findings are summarized and discussed in the concluding section \ref{Conclusion}.

\section{The method} \label{Method}

The method is based on expanding powers of the conductance or shot-noise (or any other linear statistic) in Schur functions $s_\lambda(T)$. These functions are symmetric polynomials in the transmission eigenvalues $T=\{T_1, \ldots, T_n\}$ indexed by partitions $\lambda$. In the group representation theory the Schur functions are the irreducible characters of the unitary group and hence are orthogonal. This orthogonality is quite useful since it means that the coefficients in Schur function expansions are just ``Fourier coefficients'', and, hence, can be found by integration over the unitary group. It gives an efficient way of calculating the expansion coefficients explicitly, the fact that we exploit in our approach. The Schur functions can be then averaged over the JPDF (\ref{jpd}),
\begin{equation}\label{<s>}
  \langle s_{\lambda} \rangle = \int d[T]  s_{\lambda}(T) \mathcal{P}_\beta(T)\,, \quad d[T]\equiv\prod_{i=1}^ndT_i\,,
\end{equation}
with the help of integration formulas due to Hua \cite{Hua}. The Schur function  expansions and Hua's integration formulas provide us with the necessary ingredients to compute all moments (or cumulants) of the conductance and/or shot-noise, see Sec. \ref{Moments} for the detailed analysis.

In this section, we first give a brief summary of the required facts about partitions and Schur functions, \cite{MacDonald} then develop the systematic way of performing the expansion over Schur functions and finally determine Schur function averages.

\subsection{Partitions and Schur functions}\label{Parts}

A partition is a finite sequence $\lambda=(\lambda_1, \lambda_2, \ldots , \lambda_m)$ of non-negative integers (called parts) in decreasing order  $\lambda_1 \ge \lambda_2 \ge \ldots \ge \lambda_m \ge 0$. The weight of a partition, $|\lambda|$, is the sum of its parts, $|\lambda|=\sum_j \lambda_j$, and the length, $l(\lambda)$, is the number of its non-zero parts. No distinction is made between partitions which differ only by the number of zero parts. Different partitions of weight $m$ represent different ways to write $m$ as the sum of positive integers and can be graphically visualized through the Young diagrams. For example, one has only one partition $\lambda=(1)$ in the trivial case of $m=1$; two partitions $\lambda = (2,0),(1,1)$ for $m=2$; three partitions $\lambda = (3,0,0), (2,1,0), (1,1,1)$ for $m=3$, etc.

For any partition $\lambda$ of length $l(\lambda) \le n$, one can define a symmetric polynomial $s_{\lambda}$ in $n$ variables $x_1, \ldots , x_n$ as follows:
\begin{equation}\label{schur}
s_{\lambda}(x_1, \ldots , x_n)=\frac{\det \left\{
x_i^{\lambda_j+n-j}\right\}_{i,j=1}^n}{\det\left\{
x_i^{n-j}\right\}_{i,j=1}^n}
\end{equation}
The denominator here is nothing else but the Vandermonde determinant $\Delta(x)=\prod_{i<j}(x_i-x_j)$. It divides the corresponding factor in the nominator, leaving the quotient as a homogeneous polynomial in the $x_j$'s of degree $m=|\lambda|$. These polynomials $s_{\lambda}$ are called the Schur functions. For one-part partitions, $\lambda = (r)$, Schur functions are just the complete symmetric functions, $s_{(r)}(x)=h_r$, while for partitions which have no parts other than zero or one, $\lambda =(1,\ldots, 1)\equiv (1^r)$, the Schur functions $s_{(1^r)}$ are the elementary symmetric functions $e_r(x)$. This can be verified directly from (\ref{schur}). It should be noted that the Schur functions corresponding to the partitions of $m$ form a basis in the space of homogeneous symmetric polynomials of degree $m$, so that any homogeneous symmetric polynomial can be written as a linear combination of Schur functions.

The Schur functions of matrix argument that we shall use below are defined by the rhs in (\ref{schur}) evaluated at the  eigenvalues of the matrix. Taking as an example the $n{\times}n$ matrix $T=tt^\dag$ of transmission probabilities, one has
$$
 s_{\lambda}(T) = s_{\lambda}(T_1, \ldots, T_n)\,
$$
where $T_1, \ldots, T_n$ are exactly the transmission eigenvalues that appear in (\ref{jpd}). Although not apparent from this definition, the Schur functions of matrix argument are polynomials in the matrix entries \cite{Jucobi-Trudi} and, obviously, $s_{\lambda}(T)=s_{\lambda}(XTX^{-1})$ for any non-degenerate matrix $X$.

\subsection{Schur function expansions}\label{Expansions}

In order to determine the moments of the conductance and shot noise along the lines explained above, one needs to expand the powers of these quantities  in Schur functions. To this end, it is more convenient to work with the corresponding generating functions $e^{t\sum T_j}$ or $e^{t\sum T_j(1-T_j)}$. These functions belong to the general class of multiplicative symmetric functions, where the coefficients of the Schur function expansion
\begin{equation}\label{F}
  F(x) \equiv \prod_{j} f(x_j) = \sum_{\lambda} c_{\lambda}^{(f)} s_{\lambda} (x)
\end{equation}
can be determined explicitly provided that the function $f$ is analytic in a neighborhood of $|x|=1$ in the complex $x$-plane, see, e.g., Appendix in Ref.~\cite{Fyodorov2007}. Indeed, thinking of the $x_j$'s as of the eigenvalues of a unitary matrix $U$, one can write
\begin{equation}\label{F(U)}
  F(U)=\sum_{\lambda} c_{\lambda}^{(f)} s_{\lambda}(U)\, .
\end{equation}
The main advantage of going unitary is the orthogonality of Schur functions ($d\mu(U)$ is the normalized Haar measure):
\begin{equation}\label{eq}
 \int_{U(n)}\!d\mu(U)\, s_{\lambda}(U)\, s_{\mu}^{*}(U) =\delta_{\lambda,\mu},
\end{equation}
which is a fact from the theory of group representations. One now recognizes a ``Fourier series'' in (\ref{F(U)}) and, hence,
\begin{equation}\label{clambda1}
 c_{\lambda}^{(f)} = \int_{U(n)}\!d\mu(U)\, F(U)\, s_{\lambda}^{*}(U).
\end{equation}

The integral on the rhs in (\ref{clambda1}) is a standard one in RMT. To evaluate it, one first transforms it to the eigenvalues $e^{i\theta_1},\ldots,e^{i\theta_n}$ of the unitary matrix $U$. The corresponding Jacobian is  $|\Delta(e^{i\theta})|^2$, canceling the denominator in the Schur function $s_{\lambda}^{*}(e^{i\theta})$. The resulting integral can then be evaluated with the help of the Andrejeff identity, \cite{Polya1} yielding
\begin{equation}\label{clambda2}
  c_{\lambda}^{(f)} = \det \left\{\int_{0}^{2\pi}\!\! \frac{d\theta}{2\pi} f(e^{i\theta})
  e^{-i\theta (\lambda_k-k+l)}\right\}_{k,l=1}^n.
\end{equation}
In view of the analyticity one can abandon the restriction $|x|=1$. Writing  $f(x)=\sum_j \tau_j \,{x}^j$, one brings the Schur function expansion (\ref{F}) and (\ref{clambda2}) to the following general form:
\begin{subequations}
 \label{exp}
\begin{eqnarray}
  &&\prod_{i=1}^n \biggl(\sum_{j=-\infty}^{+\infty} \tau_j\  {x_i}^j \biggr)
  = \sum_{\lambda}c_{\lambda}(\tau) s_{\lambda}(x), \\
  && c_{\lambda}(\tau) \equiv \det\bigl\{\tau_{\lambda_k-k+l}\bigr\}_{k,l=1}^n.
\end{eqnarray}
\end{subequations}
The summation here is over all partitions $\lambda$ of length $n$ or less, including empty partition $(0)$ for which $s_{\lambda}=1$.  Expansion (\ref{exp}) was also obtained by Balantekin \cite{Balantekin2000} by algebraic manipulations.

\subsection{Schur function averages}\label{Averages}

Hua in his book\cite{Hua} evaluated many useful matrix integrals. The following two are relevant in the context of our work:
\begin{equation} \label{hua2}
 \langle s_{\lambda} \rangle_{\beta=2}  \!=\! \prod_{j=1}^n\frac{\Gamma (j+1)\Gamma(\lambda_j\!+\!n\!-\!j\!+\!\alpha)}{\Gamma(\lambda_j\!+\!2n\!-\!j
 \!+\!\alpha )} \!\!\!\!\prod\limits_{1\le i < j \le n}\!\!\!(\lambda_i\!-\!\lambda_j\!-\!i\!+\!j)
\end{equation}
and
\begin{equation} \label{hua1}
 \langle s_{\lambda} \rangle_{\beta=1} \!=\! \frac{2^nn! \prod_{1\le i < j \le n} (\lambda_i-\lambda_j-i+j)}{\prod_{1\le i\le j \le n} (\lambda_i+\lambda_j+2n+2\alpha-i-j) },
\end{equation}
where the average $\langle s_{\lambda} \rangle_{\beta}$ is over the JPDF (\ref{jpd}), as in Eq.~(\ref{<s>}). If $\lambda=(0)$ then $s_{\lambda}=1$ and both integrals follow from the Selberg integral \cite{Mehta2}. It should be noted that the rhs in (\ref{hua2}) is exactly Selberg's expression and the rhs in (\ref{hua1}) can be manipulated to the one obtained by Selberg with the help of the duplication formula for the Gamma function. For non-empty partitions $\lambda$, the integral in (\ref{hua2}) is a particular case of the Kadell-Kaneko-Yan generalization \cite{Kadell1997,Kaneko1993,Yan1992} of the Selberg integral. However, the integral in (\ref{hua1}) is different, as the Kadell-Kaneko-Yan generalization of the Selberg integral for $\beta=1$ involves zonal polynomials.

\begin{table*}[t]
\renewcommand{\arraystretch}{1.75}
\caption{Schur function expansion ${\langle(\sum T_j)^m\rangle_{\beta} = m!\sum_{} c_{\lambda}\langle s_{\lambda}\rangle_{\beta}}$ for $\beta=1,2$.}
\begin{ruledtabular}
\begin{tabular}{ccccc}
 $m$ & Partition $\lambda$ & $\langle s_{\lambda}\rangle_{\beta=2}$ &
 $\langle s_{\lambda}\rangle _{\beta=1}/\langle s_{\lambda}\rangle _{\beta=2}$ & $c_{\lambda}$
 \\
\hline
\multirow{1}{*}{ 1 } & $(1)$ & ${\frac{N_1N_2}{N}}$ & ${\frac{N}{N+1}}$ & 1 \\ \hline
\multirow{2}{*}{ 2 } & $(2,0)$ & ${\frac{N_1(N_1+1)N_2(N_2+1)}{2N(N+1)}}$ & ${\frac{N+1}{N+3}}$ & $ {\frac{1}{2}}$
 \\
 & $(1,1)$ & ${\frac{N_1(N_1-1)N_2(N_2-1)}{2N(N-1)}}$ & ${\frac{N-1}{N+1}}$ & $ {\frac{1}{2}}$
 \\ \hline
\multirow{3}{*}{ 3 } & $(3,0,0)$ & ${\frac{N_1(N_1+1)(N_1+2)N_2(N_2+1)(N_2+2)}{6N(N+1)(N+2)}}$ & ${\frac{N+2}{N+5}}$ & $ {\frac{1}{6}}$
 \\
 & $(2,1,0)$ & ${\frac{(N_1-1)N_1(N_1+1)(N_2-1)N_2(N_2+1)}{3(N-1)N(N+1)}}$ & ${\frac{N}{N+3}}$ & ${\frac{1}{3}}$
 \\
  & $(1,1,1)$ & ${\frac{(N_1-2)(N_1-1)N_1(N_2-2)(N_2-1)N_2}{6(N-2)(N-1)N}}$ & ${\frac{N-2}{N+1}}$ & $ {\frac{1}{6}}$
\\  \hline
\multirow{5}{*}{ 4 } & $(4,0,0,0)$ & ${\frac{N_1(N_1+1)(N_1+2)(N_1+3)N_2(N_2+1)(N_2+2)(N_2+3)}{24N(N+1)(N+2)(N+3)}}$ & ${\frac{N+3}{N+7}}$ & ${\frac{1}{24}}$
 \\[1ex]
 & $(3,1,0,0)$ & ${\frac{(N_1-1)N_1(N_1+1)(N_1+2)(N_2-1)N_2(N_2+1)(N_2+2)}{8(N-1)N(N+1)(N+2)}}$ & ${\frac{N+1}{N+5}}$ & ${\frac{1}{8}}$
 \\
 & $(2,2,0,0)$ & ${\frac{(N_1-1)N_1^2(N_1+1)(N_2-1)N_2^2(N_2+1)}{8(N-2)(N-1)N(N+1)}}$ & ${\frac{(N-1)N^2}{(N-2)(N+2)(N+3)}}$ & $ {\frac{1}{8}}$
 \\
 & $(2,1,1,0)$ & ${\frac{(N_1-2)(N_1-1)N_1(N_1+1)(N_2-2)(N_2-1)N_2(N_2+1)}{8(N-2)(N-1)N(N+1)}}$ & ${\frac{N-1}{N+3}}$ & $ {\frac{1}{8}}$
 \\
 & $(1,1,1,1)$ & ${\frac{(N_1-3)(N_1-2)(N_1-1)N_1(N_2-3)(N_2-2)(N_2-1)N_2}{24(N-3)(N-2)(N-1)N}}$ & ${\frac{N-3}{N+1}}$ & $ {\frac{1}{24}}$
 \\ \hline
\multirow{7}{*}{ 5 } & $(5,0,0,0,0)$ & ${\frac{N_1(N_1+1)(N_1+2)(N_1+3)(N_1+4)N_2(N_2+1)(N_2+2)(N_2+3)(N_2+4)}{120N(N+1)(N+2)(N+3)(N+4)}}$ & ${\frac{N+4}{N+9}}$ & $ {\frac{1}{120}}$
 \\[1ex]
 & $(4,1,0,0,0)$ & ${\frac{(N_1-1)N_1(N_1+1)(N_1+2)(N_1+3)(N_2-1)N_2(N_2+1)(N_2+2)(N_2+3)}{30(N-1)N(N+1)(N+2)(N+3)}}$ & ${\frac{N+2}{N+7}}$ & $ {\frac{1}{30}}$
 \\
 & $(3,2,0,0,0)$ & ${\frac{(N_1-1)N_1^2(N_1+1)(N_1+2)(N_2-1)N_2^2(N_2+1)(N_2+2)}{24(N-1)N^2(N+1)(N+2)}}$ & ${\frac{(N-1)N(N+2)}{(N-2)(N+3)(N+5)}}$ & ${\frac{1}{24}}$
 \\
 & $(3,1,1,0,0)$ & ${\frac{(N_1-2)(N_1-1)N_1(N_1+1)(N_1+2)(N_2-2)(N_2-1)N_2(N_2+1)(N_2+2)}{20(N-2)(N-1)N(N+1)(N+2)}}$ & ${\frac{N}{N+5}}$ & $ {\frac{1}{20}}$
 \\
 & $(2,2,1,0,0)$ & ${\frac{(N_1-2)(N_1-1)N_1^2(N_1+1)(N_2-2)(N_2-1)N_2^2(N_2+1)}{24(N-2)(N-1)N^2(N+1)}}$ & ${\frac{(N-2)(N-1)N}{(N-3)(N+2)(N+3)}}$ & ${\frac{1}{24}}$
 \\
 & $(2,1,1,1,0)$ & ${\frac{(N_1-3)(N_1-2)(N_1-1)N_1(N_1+1)(N_2-3)(N_2-2)(N_2-1)N_2(N_2+1)}{30(N-3)(N-2)(N-1)N(N+1)}}$ & ${\frac{N-2}{N+3}}$ & $ {\frac{1}{30}}$
 \\
 & $(1,1,1,1,1)$ & ${\frac{(N_1-4)(N_1-3)(N_1-2)(N_1-1)N_1(N_2-4)(N_2-3)(N_2-2)(N_2-1)N_2}{120(N-4)(N-3)(N-2)(N-1)N}}$ & ${\frac{N-4}{N+1}}$ & $ {\frac{1}{120}}$
\end{tabular}
\end{ruledtabular}
\end{table*}

Hua's identities are valid for arbitrary  continuous $\alpha>0$. Specifying further to our case of
$\alpha=\frac{\beta}{2}(|N_1-N_2|+1)$ and $N \equiv N_1+N_2$, we arrive after a simple algebra at
\begin{eqnarray}\label{mg2a}
 \langle s_{\lambda} \rangle_{\beta=2} &=& c_{\lambda} \prod_{j=1}^{l(\lambda)} \frac{(\lambda_j+N_1-j)!}{(N_1-j)!} \frac{(\lambda_j+N_2-j)!}{(N_2-j)!}\nonumber \\
 && \ \times \frac{(N-j)!}{(\lambda_j+N-j)!},
\end{eqnarray}
where we have introduced the coefficient \cite{clambda}
\begin{equation}\label{c_lambda}
c_{\lambda} = \frac{\prod_{1\le i<j\le l(\lambda)}(\lambda_i-i-\lambda_j+j)}{\prod_{j=1}^{l(\lambda)}\, (l(\lambda)+\lambda_j-j)!},
\end{equation}
and
\begin{eqnarray}\label{mg1a}
  \langle s_{\lambda} \rangle_{\beta=1} &=& c_{\lambda} \prod_{j=1}^{l(\lambda)} \frac{(\lambda_j+N_2-j)!}{(N_2-j)!}\nonumber \\
  && \times \!\!\! \prod_{1\le i\le j\le N_2}\!\!\! \Bigl(\frac{N+1-i-j}{N+1+\lambda_i+\lambda_j-i-j}\Bigr).\quad
\end{eqnarray}
The symmetry between $N_1$ and $N_2$ is not apparent in
(\ref{mg1a}). One can rearrange the terms in the second product on
the rhs in (\ref{mg1a}) to make this symmetry apparent
\begin{eqnarray}\label{mg1aa}
 \langle s_{\lambda} \rangle_{\beta=1} &=& c_{\lambda}\ \prod_{j=1}^{l(\lambda)} \frac{(\lambda_j+N_1-j)!}{(N_1-j)!}\ \frac{(\lambda_j+N_2-j)!}{(N_2-j)!}
 \nonumber \\
 &&\times\!\prod_{1\le i\le j\le l(\lambda)} \frac{N+1-i-j}{N+1+\lambda_i+\lambda_j-i-j}
 \nonumber \\
 &&\times \prod_{i=1}^{l(\lambda)} \frac{(N-l(\lambda)-i)!}{(\lambda_i+N-l(\lambda)-i)!}\,.
\end{eqnarray}
We note that the obtained expressions for $\langle s_{\lambda}\rangle$ in terms of $N_1$ and $N_2$ yield zero if  the length of the partition $\lambda$ is greater than  $n=\min(N_1,N_2)$ so when averaging Schur function expansions one need not bother about the restriction $l(\lambda) \le n$.

The Schur function average for $\beta=2$ in terms of the channel numbers $N_1$ and $N_2$, Eq.~(\ref{mg2a}), has a simple structure, being a ratio of polynomials
\begin{equation}\label{b:e4b}
\langle s_{\lambda} \rangle_{\beta=2} = c_{\lambda}\
\prod_{j=1}^{|\lambda|}\frac{(N_1-a_j)(N_2-a_j)}{(N-a_j)}
\end{equation}
where the $a_j$'s are integers. Expression (\ref{mg1aa}) for $\langle s_{\lambda} \rangle_{\beta=1}$ is less revealing. We found it useful to have the Schur function averages tabulated, see Table I for averages corresponding to partitions of $m$, $m=1,\ldots,5$.
This table suggests that
\[
\langle s_{\lambda} \rangle_{\beta=1} = c_{\lambda}\
\prod_{j=1}^{|\lambda|}\frac{(N_1-a_j)(N_2-a_j)}{(N-b_j)},
\]
where the $a_j$ are the same as in (\ref{b:e4b}) and $b_j$'s are also integers. It would be generally desirable to understand the nature of the cancelations in (\ref{mg1aa}) and to find a rule relating $b_j$ to $\lambda$. \cite{b_j}

\section{Moments and cumulants of the \newline conductance and shot-noise}\label{Moments}

We now apply the results obtained in the previous two sections to calculate the moments of the conductance and shot-noise in a closed form. The final expressions involve summation over all partitions of $r$ in the case of conductance and $2r$ in the case of shot-noise, with $r$ being the order of the moment. The cumulants $\kappa_r$ can be obtained from the moments $\mu_r$ with the help of the well-known recursion
\begin{equation}\label{cumm}
 \kappa_{r} = \mu_{r}-\sum_{j=1}^{r-1} {r{-}1 \choose r{-}j}  \mu_{r-j} \kappa_{j}\,.
\end{equation}
This method is well suited for analytic computations of lower order cumulants and also can be straightforwardly implemented in a computer algebra system for computations of higher order cumulants symbolically. Since a number of the partitions of the given $r$ grows $\sim\exp\{\pi\sqrt{2r/3}\}/(4r\sqrt{3})$ only asymptotically at $r\to\infty$ (i.e. slower than pure exponential), our method is very efficient for computing the cumulants up to reasonably large orders, as discussed below.

\subsection{Conductance}

The moments of the conductance can be obtained from the generating function
\begin{equation}\label{F_g}
 F_g(t) = \left\langle e^{t\sum_iT_i}\right\rangle.
\end{equation}
The desired Schur function expansion for the exponential function $e^{\sum_iT_i}$ can be read from (\ref{exp}) by choosing $\tau_j=1/j!$ there. Throughout this paper we use the convention that $1/j!=0$ for $j<0$. The factorial determinant in (\ref{exp}) can be evaluated by elementary transformations on its rows or columns and the answer turns out to be exactly the coefficient $c_\lambda$ introduced in (\ref{c_lambda}). Recalling that the Schur functions are homogeneous, thus $s_{\lambda}(tT)=t^{|\lambda|}s_{\lambda}(T)$, one turns the Schur function expansion of the moment generating function into the following series in powers of $t$:
\begin{equation}\label{eschur}
 F_g(t) = \sum_{r=0}^{\infty} t^r \sum_{\lambda \vdash r}\ c_{\lambda} \langle{s_{\lambda}}\rangle\,.
\end{equation}
The second sum on the right is over all partitions of $r$. From (\ref{eschur}) one easily obtains all moments of the conductance:
\begin{equation}\label{<g^r>}
 \langle{g^r}\rangle = r!\  \sum_{\lambda \vdash r} c_{\lambda} \langle{s_{\lambda}}\rangle\,, \quad r=1,2,\ldots,.
\end{equation}

For $\beta=2$ this expression together with Eqs.~(\ref{mg2a}) and (\ref{c_lambda}) reproduces the recent result of Novaes. \cite{Novaes2008}

With Eq.~(\ref{<g^r>}) in hand, one can obtain cumulants by making use of Eq.~(\ref{cumm}). On this way we have successfully reproduced the first four cumulants which have been obtained before (exactly for any $\beta$). For the reference purpose, we state explicitly the conductance variance \cite{Brouwer1996}
\begin{equation}\label{var_g}
 \frac{\mathrm{var}(g)}{\langle g \rangle} = \left\{
 \begin{array}{ll}
  \displaystyle
  \frac{2(N_1+1)(N_2+1)}{N(N+1)(N+3)}, & \ \ \beta=1 \\[2ex]
  \displaystyle
  \frac{N_1N_2}{(N-1)N(N+1)}, & \ \ \beta=2
 \end{array} \right.,
\end{equation}
with $\langle{g}\rangle=N_1N_2/(N+\frac{2}{\beta}-1)$ being the conductance average,
and the third cumulant \cite{Savin2008}
\begin{equation}\label{skew_g}
 \frac{\langle\langle g^3\rangle\rangle}{\mathrm{var}(g)} = \left\{
 \begin{array}{ll}
  \displaystyle
  \frac{4[1-(N_1-N_2)^2]}{(N-1)(N+1)(N+5)}, & \ \ \beta=1 \\[2ex]
  \displaystyle
  -\frac{2(N_1-N_2)^2}{(N-2)N(N+2)}, & \ \ \beta=2
 \end{array} \right.,
\end{equation}
which is a measure of the skewness of the probability distribution.
An explicit expression for the fourth cumulant is quite lengthy for
arbitrary $N_{1,2}$ and the corresponding large $N$ expansion can be
found in Ref.~\cite{Savin2008} (see also below). However, in the
particular case of $N_1=N_2=n$, it can be simplified further to the
following compact form:
\begin{widetext}
\begin{equation}
 \frac{\langle\langle{g^4}\rangle\rangle}{\mathrm{var}(g)} = \left\{
 \begin{array}{ll}
  \displaystyle
  -\frac{3(4n^4+20n^3+43n^2+53 n+24)}{(n+1)(2n-1)(2n+1)^2(2n+3)(2n+5)(2n+7)}, & \ \ \beta=1 \\[2ex]
  \displaystyle
  \frac{3}{2(2n-3)(2n-1)(2n+1)(2n+3)}, & \ \ \beta=2
 \end{array} \right. .
\end{equation}
\end{widetext}

Let us now discuss higher cumulants of the  conductance, $\langle\langle g^r \rangle\rangle$. Their explicit expressions are cumbersome and we consider mainly the physically interesting cases of the small or large channel numbers. In the quantum regime of a few open channels, we have found that these cumulants do not show a pronounced decay with increasing $r$, see Fig.~1. In this case, the distribution function is strongly non-Gaussian. However, as the number of channels in the both leads increases, the system approaches the semiclassical (`metalic') regime where one should expect \cite{Altshuler1986b} the following dependence of the cumulants on the total number of channels, $N=N_1+N_2$:
\begin{equation}\label{<<g^r>>}
 \langle\langle g^r \rangle\rangle\sim\langle{g}\rangle^{2-r}\sim N^{2-r}.
\end{equation}
The same scaling is generally applicable to any linear statistic on transmission  eigenvalues (e.g., shot-noise), implying a Gaussian distribution in the limit $N\to\infty$  \cite{Politzer1989,Beenakker1993} (see, however, the next section for discussion).

\begin{figure}[t]
\centering
\includegraphics[width=0.425\textwidth]{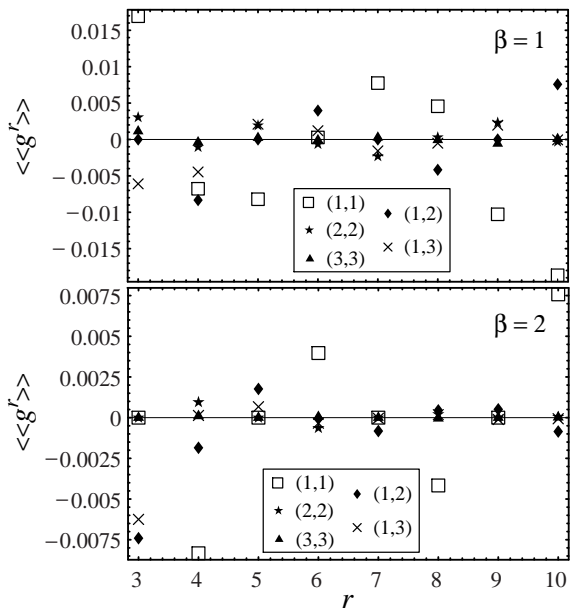}
\caption{The third to tenth cumulants of the conductance for chaotic cavities with a few open channels in the case of preserved ($\beta=1$) or broken ($\beta=2$) time-reversal symmetry. The corresponding values $(N_1,N_2)$ of the channel numbers are as indicated in the legend.}
\end{figure}

We have performed the asymptotic analysis  of our exact RMT
expressions in the limit when both $N_{1,2}\gg1$. It suggests that
the leading order term in the $1/N$ expansion of the $r$-th
cumulant, $r\ge3$, has the following general structure:
\begin{equation}\label{cumg_asym}
 \frac{\langle\langle g^r \rangle\rangle}{\langle g \rangle} \simeq \frac{(r-1)!}{(\beta/2)^{r-1}}
 \frac{ N_1N_2 (N_1-N_2)^2 }{ N^{3(r-1)} } G_{2(r-3)},
\end{equation}
where $G_{m}(N_1,N_2)$ is an  independent of $\beta$ homogeneous
symmetric polynomial of order $m$, see Table II for the first four
ones. The expression on the rhs in (\ref{cumg_asym}) is of the order
of $N^{1-r}$, being in agreement with the above estimate
(\ref{<<g^r>>}) obtained within a different approach (weak
localisation diagrammatics). The next to leading order term in the
$1/N$ expansion of $ {\langle\langle g^r \rangle\rangle}/{\langle g
\rangle}$, the so-called weak localization correction, is of the
order of $N^{-r}$. It vanishes for systems with broken time-reversal
symmetry ($\beta=2$). Further terms in this $1/N$ expansion can be
easily computed as well if necessary.

\begin{table}[b]
\caption{The first four polynomial $G_m(N_1,N_2)$, Eq.~(\ref{cumg_asym}).}
\begin{ruledtabular}
\renewcommand{\arraystretch}{1.5}
\begin{tabular}{cc}
 \ $r$ & $G_{2(r-3)}(N_1,N_2)$ \\
\hline
 \ 3 & 1 \\
 \ 4 & $N_1^2 - 4N_1N_2 + N_2^2$ \\
 \ 5 & $N_1^4 - 10N_1^3N_2 + 22N_1^2N_2^2 - 10N_1N_2^3 + N_2^4$ \\
 \ 6 & $N_1^6 - 18N_1^5N_2 + 88N_1^4N_2^2 - 150 N_1^3N_2^3 + (N_1\rightleftarrows N_2)\quad$
\end{tabular}
\end{ruledtabular}
\end{table}

In the special case of symmetric cavities, $N_1=N_2=n$, the leading term (\ref{cumg_asym}) in the $1/N$ expansion of the cumulants vanishes for all $r\ge3$ and so does the next-to-leading term of the expansion of any odd cummulant (independently of $\beta$, it contains a factor $(N_1-N_2)^2$ explicitly). This indicates that the Gaussian distribution is approached in this case much faster as compared to (\ref{<<g^r>>}) -- (\ref{cumg_asym}).

Generally, we note that in the symmetric case all odd cumulants at
$\beta=2$ must vanish identically, as it follows by the simple
symmetry argument \cite{Savin2008} (indeed, the joint distribution
(\ref{jpd}) becomes then symmetric under the change of all
$T_j\to1{-}T_j$ implying the symmetry of the conductance
distribution about its mean $\frac{n}{2}$). It has been recently
checked \cite{Novaes2008} that representation (\ref{<g^r>}) at
$\beta=2$ satisfies this property. For the even cumulants,  the
$1/n$ expansion of our exact expressions gives the following leading
term at $n\gg1$ ($k\geq2$):
\begin{equation}\label{cumg_n2}
 \langle\langle g^{2k} \rangle\rangle_{\beta=2} \simeq \frac{(2k{-}1)!}{4(4n)^{2k}}
\end{equation}
that agrees with the recent result by Osipov and  Kanzieper
\cite{Osipov2008} obtained by a completely different method. In the
case of $\beta=1$, we have found with the help of symbolic
computations in Mathematica that
\begin{equation}\label{cumg_n}
 \langle\langle g^r \rangle\rangle_{\beta=1} \simeq \frac{(r-1)!}{4(2n)^r} \times \left\{
 \begin{array}{ll}
  1, & \mbox{odd $r$} \\[1ex]
  \displaystyle \frac{-2n(r{-}3)!!}{r!!}, & \mbox{even $r$}
 \end{array}\right. .
\end{equation}
for $r=3,4,\ldots, 16$. Correspondingly, we put forward the
conjecture that Eq.~(\ref{cumg_n}) holds for all $r\ge 3$.

\subsection{Shot-noise}

Having an aim to find also the joint moments of the conductance and shot-noise, we consider the generating function for the moments of $ag + p$:
\begin{equation}\label{F(t,a)}
 F(t,a) \equiv \left\langle e^{t(ag+p)} \right\rangle = \left\langle \prod_{i=1}^ne^{t(a+1)T_i+tT_i^2} \right\rangle\,. \
\end{equation}
The moment generating function of shot-noise is then simply given by $F_{p}(t)= F(t,0)$ whereas that of the conductance follows as $F_{g}(t)= \lim_{a\to\infty}F(t/a,a)$. At finite $a$, the quantity $ag+p$ has a physical meaning of the total noise including both thermal and shot-noise contributions, with $a$ being then the known function of the temperature and applied voltage.\cite{Blanter2000}

The exponential function in (\ref{F(t,a)}) can be expanded in Schur functions $s_{\lambda}(T)$ with the help of the general identity (\ref{exp}). On multiplying two exponential series, one obtains
\begin{subequations}\label{pexp}
\begin{eqnarray}
 && F(t,a) = \sum_{\lambda}c_{\lambda}(t,a) \langle s_{\lambda}\rangle, \\
 && c_{\lambda}(t,a) = \det \bigl\{\pi_{\lambda_i-i+j}(t,a) \bigr\}_{i,j=1}^n ,
\end{eqnarray}
\end{subequations}
where $\pi_r(t,a)$ are polynomials in $t$ and $(a+1)$,
\begin{eqnarray}\label{pi}
 \pi_r(t,a)=\sum_{k=0}^{\lfloor r/2 \rfloor}\frac{(-1)^k (a+1)^{r-2k}\ t^{r-k}}{k!\ (r-2k)!}.
\end{eqnarray}

In order to extract from this the moments of $ag+p$ one needs to expand the coefficients $c_{\lambda}(t,a)$ in powers of $t$. After some algebra, see Appendix \ref{App1} for details, one arrives at the desired expansion
\begin{equation}\label{F(t,a)exp}
 F(t,a) = \sum_{r=0}^{\infty}t^r \sum_{m=0}^{r}(-1)^m (1{+}a)^{r-m} \!\!\sum_{\lambda \vdash r+m} f_{\lambda,m} \langle s_{\lambda} \rangle,
\end{equation}
where
\begin{equation}\label{flambda}
 f_{\lambda,m} = \sum_{k_1+\ldots +k_{l(\lambda)}=m}
 \det \Bigl\{ \frac {1} {{k_i}!(\lambda_i-i+j-2k_i)!} \Bigr\}.
\end{equation}
The determinant on the rhs (\ref{flambda}) can be evaluated in terms of the partition $\lambda$ leading to an explicit expression for the coefficients $f_{\lambda,m}$, see Eq.~(\ref{fexp}). In the particular case of $m=0$, $f_{\lambda,0}$ is just the coefficient $c_{\lambda}$ given by (\ref{c_lambda}). We note that $f_{\lambda,m}$ depend only on $\lambda$ and $m$ and not on $n$. The summation indices $k_j$ in (\ref{flambda}) run over all integers from 0 to $m$ and are not subject to any ordering.  From expansion (\ref{F(t,a)exp}), one easily finds that the $r$-th moment of the total noise reads as follows:
\begin{eqnarray}\label{gpexp}
 \langle(ag + p)^r \rangle &=&  r! \sum_{m=0}^{r} (-1)^m (1{+}a)^{r-m} \nonumber \\
 && \times \sum_{\lambda \vdash r+m} f_{\lambda,m} \langle s_{\lambda} \rangle,
\end{eqnarray}
where the second sum is over all partitions of $r{+}m$. The joint moment  $\langle g^k p^{r-k}\rangle$ of the conductance and shot-noise is then given by Eq.~(\ref{gpexp}), with $(1{+}a)^{r-m}$ being replaced by the binomial coefficient $r-m \choose k$. It is interesting to note that by setting $a=-1$, one also obtains the moments of the sum of squares of the transmission coefficients:
$$
 \bigl\langle\Bigl(\sum_i {T_i}^2\Bigr)^{r}\bigr\rangle = r! \sum_{\lambda\vdash 2r} f_{\lambda,r} \langle s_{\lambda} \rangle.
$$
To the best of our knowledge the above formulas have not been reported in the literature before.

We now focus on the analysis of the shot-noise cumulants. Expansion (\ref{gpexp}) successfully reproduces the general $\beta$ results for the shot-noise average \cite{Savin2006}
\begin{equation}
 \langle p \rangle = N_1N_2\frac{\beta}{2}\frac{\mathrm{var}(g)}{\langle g \rangle}
\end{equation}
and for the shot-noise variance\cite{Savin2008}. The explicit expression for the later is rather lengthy (see Ref.~\cite{Savin2008} for the corresponding large $N$ expansion) but turns out to be quite compact in the particular case of $N_1=N_2$=n:
\begin{widetext}
\begin{equation}
 \frac{\mathrm{var}(p)}{\langle p \rangle} = \left\{
 \begin{array}{ll}
  \displaystyle
  \frac{8n^5+60n^4+142n^3+91n^2-49n-36)}{2(n+1)(2n-1)(2n+1)(2n+3)(2n+5)(2n+7)}, & \ \ \beta=1, \\[2ex]
  \displaystyle
  \frac{4n^4-9n^2+3}{4n(2n-3)(2n-1)(2n+1)(2n+3)}, & \ \ \beta=2
 \end{array} \right. .
\end{equation}
\end{widetext}

Higher cumulants of shot-noise, $\langle\langle{p^r}\rangle\rangle$, similarly to those of conductance, are non-vanishing when the number of channels is small, implying a strongly non-Gaussian distribution also in this case. In the opposite limit of the  large number of channels, $N_{1,2}\gg1$, we have found the leading term of the $1/N$ expansion to have the following structure:
\begin{eqnarray}\label{cump_asym}
 \frac{\langle\langle p^r \rangle\rangle}{\langle p \rangle} &=& \frac{(r-1)!}{(\beta/2)^{r-1}}
 \frac{ (N_1-N_2)^2 }{ N^{5(r-1)} }   \nonumber \\
 && \times (N_1^2-4N_1N_2+N_2^2)^3 P_{4(r-3)},
\end{eqnarray}
with $P_m(N_1,N_2)$ being an independent of $\beta$ homogeneous polynomial of order $m$, see Table III. The next order term of the expansion has been found to have similar structure to that of the conductance (explicit expressions being, of course, different), thus the same conclusions apply for this term, too.

\begin{table}[b]
\caption{The first four polynomial $P_m(N_1,N_2)$, Eq.~(\ref{cump_asym}).}
\renewcommand{\arraystretch}{1.5}
\begin{ruledtabular}
\begin{tabular}{cc}
 $r$ & $P_{4(r-3)}(N_1,N_2)$ \\
\hline
 3 & 1 \\
 4 & $N_1^4 - 16 N_1^3N_2 + 34 N_1^2N_2^2 - 16 N_1N_2^3 + N_2^4$ \qquad\ \\
 5 & $N_1^8 - 38 N_1^7N_2 + 385 N_1^6N_2^2 - 1344 N_1^5N_2^3 + 2008 N_1^4N_2^4$ \\
   & $ + (N_1\rightleftarrows N_2)$ \\
 6 & $N_1^{12} - 66 N_1^{11}N_2 + 1345 N_1^{10}N_2^2 - 11680 N_1^9N_2^3 + 49699 N_1^8N_2^4$ \\
   & $ - 114598 N_1^7N_2^5 + 150662 N_1^6N_2^6 + (N_1\rightleftarrows N_2)$
\end{tabular}
\end{ruledtabular}
\end{table}

We consider now the particular case of $N_1=N_2=n$. In contrast to the conductance, both even and odd cumulants of shot-noise are non-vanishing at finite $n$, even for systems with broken time reversal symmetry ($\beta=2$).  In the limit of $n\gg1$, the $1/n$ expansion of our exact expressions suggests the following asymptotic behavior of the $r$-th cumulant of the shot-noise ($r\geq3$):
\begin{equation}\label{cump_n2}
 \langle\langle p^r \rangle\rangle_{\beta=2} \simeq \frac{(r-1)!}{4(8n)^r}
\end{equation}
and
\begin{equation}\label{cump_n1}
 \langle\langle p^r \rangle\rangle_{\beta=1} \simeq \frac{(r-1)!}{8(4n)^r} \times \left\{
 \begin{array}{ll}
  1, & \mbox{odd $r$} \\[1ex]
  \displaystyle \frac{-4n(r{-}3)!!}{r!!}, & \mbox{even $r$}
 \end{array}\right. .
\end{equation}
Equation (\ref{cump_n2}) agrees with the very recent result \cite{Osipov2009} obtained by a different method. We have been able to verify by symbolic computations in Mathematica that Eq.~(\ref{cump_n1}) holds up to the 8-th cumulant and, thus, conjecture it to hold for any $r\geq3$.

\section{Distribution functions}\label{Distributions}

We consider now the distribution function of the conductance,
$P_g^{(\beta)}(x) = \langle\delta(x-g)\rangle $, and that of shot-noise,
$P_p^{(\beta)}(x) = \langle\delta(x-p)\rangle $, with $g$ and $p$ being defined in (\ref{g,p}). Explicit expressions for the conductance distribution can be found in the particular cases of $n=1,2$. At $N_1=1$ and $N_2=K\geq1$, Eq.~(\ref{jpd}) readily gives  $P_{g,n=1}^{(\beta)}(g)=(\beta K/2)g^{\beta K/2-1}$ for $0<g<1$, and zero otherwise. In the case of $N_1=2$ and arbitrary $N_2=K\ge2$, the conductance distribution can also be found by performing integrations that feature in the definition (see (\ref{pg2int}) in Appendix \ref{App2}), with the final result being
\begin{eqnarray}\label{pg2}
 P_{g,n=2}^{(\beta)}(g) &=& Kg^{\beta K-1} \bigl[ X_1 - (-1)^{(\beta K-1)/2} X_2\, \Theta(g-1)
 \nonumber \\
 && \!\!\times\! \sum_{j=0}^{\beta} {\beta \choose j} \textstyle B_{1-g} \bigl( \frac{\beta}{2}(K{-}1)+j, 1-\beta K \bigr) \bigr]\qquad
\end{eqnarray}
for $0<g<2$ and zero otherwise, see Fig.~2. Here $B_{z}(a,b)$ is the incomplete beta-function,  $\Theta(x)$ stands for the Heaviside step function, and the constants $X_{1,2}$ are given by
$X_1 = \frac{\Gamma[\beta(K+1)/2+1] \Gamma(\beta K/2) }{ \Gamma(\beta/2)\Gamma(\beta K)}$ and
$X_2 = \frac{\Gamma[\beta(K+1)/2+1] }{ \Gamma(\beta)\Gamma[\beta(K-1)/2]}$. Expression (\ref{pg2}) holds for arbitrary positive integer $\beta$. In the particular case of $\beta=1$, it can be simplified further, yielding $P_{g,n=2}^{(1)}(g) = \frac{1}{2}K(K{+}1)[(\frac{g}{2})^{K-1} - (g{-}1)^{(K-1)/2} \Theta(g{-}1)]$, in agreement with Ref.~\cite{Garcia-Martin2001}.

\begin{figure}[t]
 \includegraphics[width=0.45\textwidth]{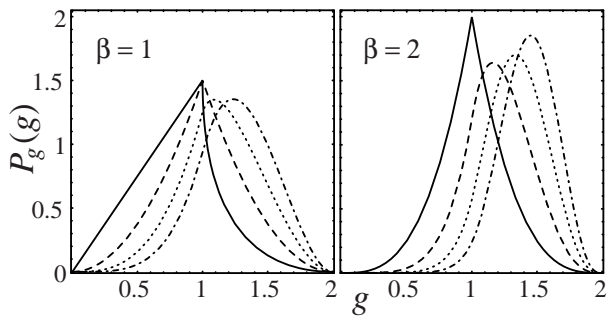}
 \caption{\label{fig2} The conductance distribution (\ref{pg2}) for chaotic cavities with preserved ($\beta=1$) or broken ($\beta=2$) time-reversal symmetry. The number of channels is fixed to $N_1=2$ in one lead and varied in the other, $N_2=2,3,4$ or 5 (solid, dashed, dotted or dash-dotted lines, respectively). Non-analyticity of the distribution (a cusp point at $g=1$) becomes less pronounced as $N_2$ increases.}
\end{figure}

It is possible to find explicit expressions for the distribution function of the conductance beyond the cases discussed above. However, the final answers become more cumbersome as the channel numbers grow, thus being almost of little practical use. For the shot-noise distribution, the situation is not satisfactory even for small channel numbers:  we are not aware of explicit results for the shot-noise distribution except for the simplest case of $N_{1,2}=1$. \cite{Pedersen1998}. One practical way to solve this problem is to construct approximations to the distribution functions in terms of cumulants by making use of the Edgeworth expansion. \cite{Cramer} It turns out that such approximations are fairly accurate in the bulk of distribution even for small channel numbers.

In the limit $N_{1,2}\gg 1$ the conductance and shot-noise distributions follow the Gaussian law
\begin{equation}\label{gauss}
 \phi_0(x)=\frac{1}{\sqrt{2\pi \sigma^2}} e^{-\frac{(x-\mu)^2}{2 \sigma^2 }}
\end{equation}
where $\mu$ and $\sigma^2$ are the corresponding mean value and variance, respectively. The Edgeworth expansion is a $1/N$ expansion around the Gaussian law. Denoting the $r$-th cumulant by  $\kappa_r$, the first correction to the Gaussian law is given by
$\phi_1(x)=-\frac{1}{3!}\kappa_3\partial^3\phi_0(x) $ and the next one is given by $\phi_2(x)=(\frac{1}{4!}\kappa_4 \partial^4 + \frac{10}{6!}\kappa_3^2 \partial^6)\phi_0(x)$. Higher order corrections involve higher order cumulants\cite{Blinnikov1998}. Restricting ourselves to the first four cumulants, we get the following approximation to the distribution functions of interest:
\begin{equation}\label{EW}
 P_{g,p}^{(\beta)}(x) \simeq \sum_{k=0}^2 \phi_k(x) .
\end{equation}
The advantage of the Edgworth expansion is that it is a true asymptotic series, with the controlled error (e.g., in our case the error of approximation (\ref{EW}) is estimated to be of the order of $\frac{1}{N^3}$). With the higher order cumulants being readily available from Eqs. (\ref{<g^r>}), (\ref{gpexp}) and (\ref{cumm}), one can easily improve the accuracy of the approximation by adding higher order corrections if needed.

\begin{figure}[t]
 \includegraphics[width=0.45\textwidth]{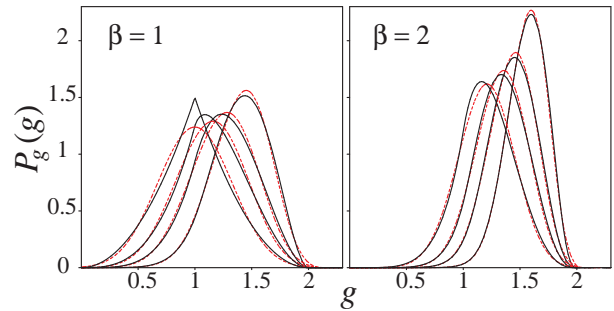}
 \caption{\label{fig3} Comparison of the exact conductance distribution (Eq.~(\ref{pg2}), black solid lines) with the Edgeworth approximation (Eq.~(\ref{EW}), red dashed lines) for chaotic cavities with preserved ($\beta=1$) or broken ($\beta=2$) time-reversal symmetry. The number of channels is fixed to $N_1=2$ in one lead and varied in the other, $N_2=3,4,5,7$ (lines from left to right, respectively). The purely Gaussian approximation would produce much stronger systematic deviations, thus being not sufficient at small channel numbers at all.}
\end{figure}

We found it instructive to compare the above approximation (\ref{EW}) with the exact results which are available for the conductance distribution. To our surprise, we found the Edgeworth approximation to be fairly accurate already for $N_1=2$ and $N_2\gtrsim 4$, see Fig.~\ref{fig3}. The agreement between the approximate and exact distributions gets even better for $N_{1,2}\ge 3$.

The Edgeworth approximation fails near the edges and also if the distribution has strong singularities. Therefore, it is tempting to look for an alternative and exact representation for the distribution functions of the conductance and shot-noise. We note that at any finite number of channels each of these distributions  has a finite support, namely, $0\le g \le n\equiv I_g$ and $0\le p \le \frac{n}{4}\equiv I_p$, being identically zero outside this region. It is, therefore, natural to represent the distribution functions in the following form (henceforth, variable and index $x = g, p$):
\begin{equation}\label{P(x)}
  P_x^{(\beta)}(x) = \sum_{m=1}^{\infty}\frac{2}{I_x}\sin\left(\frac{m\pi x}{I_x}\right)
  C^{(\beta)}_x(m)\,,
\end{equation}
as the Fourier series over the interval of support, cf. Schur function expansion (\ref{F(U)}).

We now show that the Fourier coefficients are given by Pfaffians \cite{pfaff} as follows:
\begin{equation}\label{C(m)}
 C^{(\beta)}_x(m) = \frac{n!}{\mathcal{N}_\beta}\, \mathrm{Im\,Pfaff}[A^{(\beta)}_x(m)].
\end{equation}
Before establishing explicit forms of the anti-symmetric matrices $A^{(\beta)}_x(m)$, it is useful to note the following. The Fourier coefficients turn out to decay generally as $C^{(\beta)}_x(m)\sim m^{-\nu_x}$ as $m\to\infty$, where exponent $\nu_x>1$ depends on the case considered. This readily leads to the observation that the everywhere continuous distribution function (\ref{P(x)}) contains (rather weak) singularities at integer points of division of the support interval $[0,I_x]$, as certain derivatives become discontinuous at these points (see further Ref.~\cite{Sommers2007a} for an alternative geometric interpretation of these singularities). Such a non-analyticity is expected to become less and less pronounced when the number of channels grows, as the bulk of the distribution is described then by a Gaussian law. However, the asymptotic behavior near the edges of support is always characterized by a power law \cite{Sommers2007a}. Consequently, even at $n\gg1$ the distribution remains to be weakly singular at the junction of Gaussian and power-law regimes \cite{Vivo2008}, see also Ref.~\cite{Osipov2008}.

We start the derivation of (\ref{C(m)}) with the simplest case of unitary symmetry, $\beta = 2$, and consider first the conductance distribution
\begin{equation*}
 P_g^{(2)}(g) = \frac{1}{\mathcal{N}_2}\int \!d[T] \prod_{j=1}^n  T_j^{\alpha-1} \Delta(T)^2 \delta \Bigl(g - \sum_{i=1}^n T_i\Bigr).
\end{equation*}
Writing $\Delta(T) = \det\{T_j^{i-1}\}$, with $i,j = 1,\ldots, n$, and substituting the Fourier representation of the $\delta$-function, $\delta(g-\sum_i T_i) = \!\int\!\frac{d\omega}{2\pi} e^{-i\omega g + i\omega \sum_i T_i}$, one can interchange the order of integrations and then apply the Andrejeff identity to perform the integrations over $T_j$'s. This readily yields
\begin{subequations}
\begin{eqnarray}
 \label{P(g)gue}
 &&P_g^{(2)}(g) = \frac{n!}{\mathcal{N}_2} \!\int_{-\infty}^{\infty}\!\frac{d\omega}{2 \pi}
 e^{-i \omega g} \det [\mathcal{A}^{(2)}_g(\omega)], \\
 \label{Ague}
 &&\bigl[\mathcal{A}^{(2)}_g(\omega)\bigr]_{kl} = \int_0^1\!dT T^{\alpha+k+l-3} e^{i \omega T}.
\end{eqnarray}
\end{subequations}
Since we know that the conductance distribution has a support only in $[0,n]$ it is actually more convenient to expand $P_g^{(2)}(g)$ in functions $\sqrt{2/n}\sin(m\pi g/n)$, with $m = 1,2,\ldots$, which form a complete and orthonormalized set on this interval. Furthermore, due to factorization of the exponential function, one has to express first everything in terms of $\exp(i m \pi T_j /n)$ and then take the imaginary part. As a result, we arrive at the final answer cast in the form of (\ref{C(m)}), where
\begin{equation}\label{A2}
 \mathrm{Pfaff}[A^{(2)}_g(m)] \equiv \det [\tilde{A}^{(2)}_g(m)]
\end{equation}
and $\tilde{A}^{(2)}_g(m) \equiv \mathcal{A}^{(2)}_g(\frac{m\pi}{n})$ is the discrete analogue of (\ref{Ague}).

In the case of orthogonal symmetry, $\beta=1$, the derivation goes along the same lines but has to be done separately for even or odd $n$. In the case of even $n$, we first consider the $T_j$'s in a special order $T_1 < T_2< \ldots < T_n$ (hence $n!$). Then it is useful to represent the Vandermonde determinant as a Gaussian integral over two kinds of Grassmann variables and integrate out further one set of them using the method of alternating variables \cite{Mehta2}, see also Ref.~$\cite{Sommers2008}$ for relevant details. The resulting expression acquires then the symplectic (antisymmetric) structure automatically and the remaining average yields (\ref{P(x)}) and (\ref{C(m)}), with
\begin{eqnarray}\label{A1even}
 \left[A^{(1)}_g(m)\right]^{\mathrm{even}\,n}_{kl} &=& \int_0^1\!dT_1\!\int_0^1\!dT_2\,\mbox{sign}(T_2-T_1)
 T_1^{\alpha+k-2} \nonumber \\ &&\times T_2^{\alpha+l-2} e^{i (m\pi/n) (T_1+T_2)}.
\end{eqnarray}
In the case of odd $n$, one has to increase artificially the number of the Grasmannians by one and proceed as before, with the final result being
\begin{equation}\label{A1odd}
 \left[A^{(1)}_g(m)\right]^{\mathrm{odd}\,n}_{kl} =
 \left[ \begin{array}{cc}
  \tilde{A}^{(1)}_{kl} & B_{k,n+1} \\ -B_{n+1,l} & 0
 \end{array} \right] .
\end{equation}
Here the $n{\times}n$ matrix $\tilde{A}^{(1)}$ is given by Eq.~(\ref{A1even}), and the $n$-dimensional vector $B$ is
\begin{equation}
  B_{k,n+1} = \int_0^1\! dT T^{\alpha+k-2}e^{i(m\pi/n)T} = B_{n+1,k}\,.
\end{equation}

For the sake of completeness, we also state the result in the case of symplectic symmetry, $\beta=4$:
\begin{equation}\label{A4}
 \left[A^{(4)}_{g}(m)\right]_{kl} = (l-k)\!\int_0^1\!dT\,T^{\alpha+k+l-4} e^{i (m\pi/n) T}\,,
\end{equation}
where now $k,l=1,\ldots,2n$. It can be obtained by reducing the fourth power of the Vandermonde determinant to the calculation of a Vandermonde determinant with the doubled dimension $2n$ subject to the additional $\delta$-function constraints for the corresponding pairs of eigenvalues. The rest is as in the orthogonal case above.

We note that it turns out to be possible to find explicit expressions for the conductance distribution in the case of $n=3$ and 4 by evaluating the above Pfaffians analytically and performing the corresponding Fourier transformation exactly.

The expressions derived in this section are also well suited for numerical calculations. It is important to note in this respect that, as the Pfaffian is defined as an analytic square root of the determinant of an antysimmetric matrix $A$, a special care has to be taken to decide for the complex determinant which sign has to be chosen. To overcome this difficulty, we outline the following general procedure.  First, we multiply the matrix $A$ with the symplectic unit $Z\equiv\mathrm{diag}\{-i\sigma_2,\ldots,-i\sigma_2\}$, $\sigma_2$ being the Pauli matrix. Then we note that the matrix $Z A$ is selfdual,  $Z A= A Z$, implying that its eigenvalues come in pairs. Taking the product of all eigenvalues of $Z A$ only once (this is exactly the quaternion determinant of $Z A$), we obtain finally the Pfaffian of $A$. As an illustration of this procedure, let us consider the simplest example of calculating the Pfaffian of ${\phantom{-}0\ \ a\ \choose -a\ \ 0\ } =i\sigma_2a$ which is equal to $a$. The matrix $(-i\sigma_2)(i\sigma_2a)=\mathrm{diag}\{a,a\}$ has obviously eigenvalues $(a,a)$, so that taking $a$ once yields the Pfaffian $a$. This is exactly the way how Pfaffians can be easily computed numerically.

Generally, one can derive the distribution function of any linear statistic on $T$, $x = \sum_j f_x(T_j)$ with a given $f_x(T)$, in complete analogy with the above lines. It yields representation (\ref{P(x)}), where matrices $A^{(\beta)}_x(m)$ are given by the above expressions (\ref{A2})\,--\,(\ref{A4}) in which all the factors $\exp\{i(m\pi/n)T\}$ have to be obviously substituted with $\exp\{i(m\pi/I_x)f_x(T)\}$, with $I_x$ being the length of the corresponding support interval. In particular, for the case of shot-noise it amounts to substituting there with $\exp\{i(4m\pi/n)T(1{-}T)\}$.

\section{Asymptotics}\label{Asymptotic}

The asymptotic behavior of the distribution functions near the edges is characterized by a power law dependence
\begin{equation}\label{asym}
 P^{(\beta)}_x(x) \simeq \left\{\begin{array}{ll}
      L_x x^{\ell_x},   &   x \to 0 \\[1ex]
      R_x (I_x-x)^{r_x},\  &   x\to I_x
  \end{array}\right. ,
\end{equation}
where both the exponents and the pre-factors can be determined exactly at arbitrary $N_{1,2}$ and any $\beta$, as shown below.

In the case of the conductance, the exponents $\ell_g$ and $r_g$ have been already reported previously \cite{Sommers2007a}, being given by
\begin{equation}\label{lr_g}
 \begin{array}{l}
  \ell_g = \alpha n + \frac{\beta}{2}(n-1)n - 1 \\[1ex]
  r_g = (n-1)(1 + \frac{\beta}{2} n)
 \end{array} .
\end{equation}
To determine the constant $L_g$, we consider $P_g(g)$ at $g<1$. In this case, the upper limit of the integrations over $T_i$'s in $P_g(g)=\langle\delta(g-\sum_{i}T_i)\rangle$ may be replaced with $g$ (due to the $\delta$-function). Scaling further all $T_i\to g T_i$ and calculating the powers of $g$ there, one readily gets $P_g(g)=L_g g^{\ell_g}$, with $L_g=\langle\delta(1-\sum T_i)\rangle$. It is worth emphasizing that this is the exact expression of the conductance distribution at $0<g<1$. \cite{Sommers2007a} The integral for $L_g$ can be then calculated by the standard RMT methods, see Appendix \ref{App2}, and is given by
\begin{equation}\label{L_g}
 L_g = \frac{1}{\Gamma(\ell_g+1)} \prod_{j=0}^{n-1}
 \frac{\Gamma[1 + \alpha + \frac{\beta}{2}(n{+}j{-}1)]}{\Gamma(1 + \frac{\beta}{2}j)}.
\end{equation}
The behavior of the distribution near the right edge can be analyzed in a similar way, yielding (\ref{asym}) with
\begin{equation}\label{R_g}
 R_g = \frac{1}{\Gamma(r_g+1)} \prod_{j=0}^{n-1}
 \frac{\Gamma[1 + \alpha + \frac{\beta}{2}(n{+}j{-}1)]}{\Gamma(\alpha + \frac{\beta}{2}j)}.
\end{equation}

\begin{figure}[t]
\includegraphics[width=0.475\textwidth]{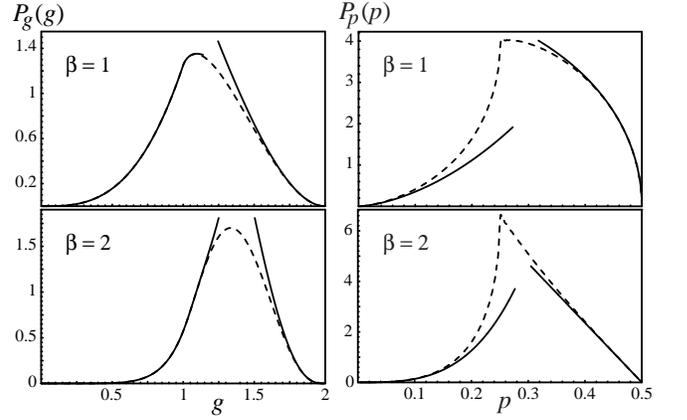}
\caption{\label{fig4} The distribution of the conductance ($P_g(g)$, left) and that of shot-noise ($P_p(p)$, right) for chaotic cavities with preserved ($\beta=1$) or broken ($\beta=2$) time-reversal symmetry. The number of channels are $N_1=2$ and $N_2=4$. Dashed lines correspond to the exact distributions whereas solid lines show the corresponding asymptotic behavior near the edges, see the text for details.}
\end{figure}

In the case of shot-noise, the corresponding exponents are found to be as follows:
\begin{equation}\label{lr_p}
 \begin{array}{l}
  \ell_p = \alpha n + \frac{\beta}{4}[(n-1)^2-\sigma] - 1 \\[1ex]
  r_p = \frac{n}{2} +\frac{\beta}{4}(n-1)n-1
 \end{array} ,
\end{equation}
where $\sigma=1$ or 0 for $n$ even or odd, respectively. The above expression for $r_g$ was already known \cite{Sommers2007a} whereas that for $\ell_p$ is new, being in agreement with the recent large $n$ result \cite{Vivo2008}. The corresponding constants can also be found exactly. We refer to Appendix \ref{App2} for further details, including the discussion of correction factors to the leading asymptotics given above.

As an illustration of the obtained results, Figure 4 shows a comparison between the exact and asymptotic behavior of the conductance and shot-noise distributions in the case of chaotic cavities with $N_1=2$ and $N_2=4$ channels.

\section{Conclusions}\label{Conclusion}

In this work, we have presented a systematic study of statistics of the conductance and shot-noise in chaotic cavities. Our approach is based on expanding symmetric functions in the transmission eigenvalues, of which the conductance and shot-noise are two examples, in Schur functions and then applying a generalization of Selberg's integral due to Hua to evaluate the averages. This leads to explicit formulas for the conductance and shot-noise cumulants in terms of the channel numbers for chaotic cavities with preserved ($\beta=1$) or broken ($\beta=2$) time-reversal symmetry.

For lower order cumulants our formulas reproduce the previously known exact results. We have performed an asymptotic analysis of the cumulants in the regime when the number of channels in both leads increases. It suggests that, generically, the $r$-th cumulant  decays as $1/N^{2-r}$ for $r\ge 3$ with the leading term containing $(N_1-N_2)^2$ as a factor; see Eqs.~(\ref{cumg_asym}) and (\ref{cump_asym}). This implies that the convergence to the limiting Gaussian law is faster in the case of symmetric cavities, $N_1=N_2=n$. In this case, we have been able to analyze the first 16 cumulants of the conductance and the first 8 cumulants of shot-noise in the limit $n\gg 1$, obtaining the leading order term in the $1/n$ expansion explicitly in terms of the cumulant order, $r$, and the channel number, $n$. For the systems with broken time-reversal symmetry, the $r$-th cumulant decays as $1/n^r$ for both conductance and shot-noise, with all the odd cumulants of the conductance being identically zero. Our results in this case, Eqs.~(\ref{cumg_n2}) and (\ref{cump_n2}), agree with those of Osipov and Kanzieper\cite{Osipov2008,Osipov2009} obtained recently by a completely different method. For systems with preserved time-reversal symmetry, we have found that the $r$-th cumulant decays as $1/n^r$ for odd $r$ and $1/n^{r-1}$ for even $r$, $r\ge 3$, for both the conductance and shot-noise. This staircase effect in the rate of the cumulant decay seems to be a novel feature which has not been reported in the literature before. One of its apparent consequences is that the convergence to the limiting Gaussian law is slower for systems with preserved time-reversal symmetry. We have also put forward our explicit formulas (\ref{cumg_n}) and (\ref{cump_n1}) for the higher order cumulants in the whole range of $r\ge 3$ as a conjecture. Proving this conjecture seems to us an interesting and challenging open problem.

As mentioned above, in the limit when the number of open channels in both leads increases, the conductance and shot-noise distributions are described by the Gaussian law. With higher order cumulants in hand, one can easily obtain next order corrections to the Gaussian law by making use of the Edgeworth expansion. We have found that such approximations to the distribution function are fairly accurate in the bulk even for small channel numbers. We have also obtained an alternative and exact representation for the distribution functions, in terms of Pfaffians, which is suitable in the whole range of support, including the edges where the distributions have a power-law dependence. Such an asymptotic behavior have been investigated in detail, the powers and corresponding pre-factors being determined exactly at any $\beta$ and $N_{1,2}$.

\begin{acknowledgments}
The authors thank W. Wieczorek for his help with numerical calculations used to produce plots on Fig.~\ref{fig4} . The two of us (B.A.K. and D.V.S.) would like to acknowledge the hospitality of the Isaac Newton Institute (Cambridge, UK) during their stay there at the programme ``Mathematics and Physics of Anderson localization: 50 Years After'', where this work has been initiated and partly completed. Financial support by the grant SFB/TR12 of DFG (H.-J.S.) and by BRIEF Award (D.V.S.) is acknowledged with thanks.
\end{acknowledgments}

\appendix

\begin{widetext}

\section{Schur function expansion, Eq.~(\ref{F(t,a)})}\label{App1}

In this appendix we expand the coefficients $c_{\lambda}(t,a)=\det \left\{\pi_{\lambda_i-i+j}(t,a) \right\}$ in powers of $t$. From (\ref{pi}) and  the definition of determinant,
\begin{eqnarray*}
 \det \left\{\pi_{\lambda_i-i+j}(t,a) \right\} &\!\!=\!\!&  \!\!\sum_{\sigma \in S_n} (-1)^{|\sigma|} \prod_{i=1}^n \sum_{k_i} \frac {(-1)^{k_i}(1+a)^{\lambda_i-i+\sigma(i)-2k_i}\ t^{\lambda_i-i+\sigma(i)-k_i}} {{k_i}!(\lambda_i-i+\sigma(i)-2k_i)!} \\ &\!\!=\!\!& \!\!\sum_{\sigma \in S_n} \!(-1)^{|\sigma|}\!\! \sum_{k_1, \ldots, k_n} \!(-1)^{K}(1+a)^{|\lambda| -2K}\ t^{|\lambda| - K}\! \prod_{i=1}^n \frac {1} {{k_i}!(\lambda_i-i+\sigma(i)-2k_i)!},
\end{eqnarray*}
where $K=\sum_i k_i$ and the first sum is over all permutations $\sigma$ in the symmetric group $S_n$. Changing the order of summations again, one can fold the sum over permutations into a determinant again. This yields
\begin{equation}\label{sumk}
 c_{\lambda}(t,a)= \sum_{m=0}^{\lfloor |\lambda|/2 \rfloor} (-1)^{m}(1+a)^{|\lambda| -2m}\ t^{|\lambda| - m}\!\! \sum_{k_1+\ldots +k_n=m}  \det \left\{ \frac {1} {{k_i}!(\lambda_i-i+j-2k_i)!} \right\}_{i,j=1}^n.
\end{equation}
The second sum here is exactly the coefficient $f_{\lambda,m}$ that appears in Eq.~(\ref{F(t,a)}). The determinant in (\ref{sumk}) can be evaluated by elementary transformations on its rows and columns,
$$
  \det \left\{ \frac {1} {{k_i}!(\lambda_i-i+j-2k_i)!} \right\}_{i,j=1}^n = \frac{ \prod_{1\le i<j\le n} (\lambda_i-i-\lambda_j+j-2k_i+2k_j) }{ \prod_{j=1}^n {k_j}!\,(n+\lambda_j-j-2k_j)!}\,.
$$
Note that for any $n\geq l(\lambda)$, where $l(\lambda)$ is the length of $\lambda$, one can safely replace $n$ in the above expressions by $l(\lambda)$. Hence, the sum over the $n$-tuples $(k_1,\ldots,k_n)$ of integers in (\ref{sumk}) can be replaced by the sum over the $l(\lambda)$-tuples $(k_1, \ldots, k_{l(\lambda)})$, yielding
\begin{equation}\label{fexp}
  f_{\lambda,m}  = \sum_{k_1+\ldots +k_{l(\lambda)}=m} \frac{\prod_{1\le i<j\le l(\lambda)} (\lambda_i-i-\lambda_j+j-2k_i+2k_j) }{ \prod_{j=1}^{l(\lambda)} {k_j}!\, (l(\lambda)+\lambda_j-j-2k_j)!}.
\end{equation}
We now substitute the obtained expression for $c_{\lambda}(t,a)$ back in (\ref{pexp}) to obtain
$$
  F(t,a)= \sum_{r=0}^{\infty}\sum_{m=0}^{\lfloor r/2 \rfloor} (-1)^m
  (1+a)^{r-2m} t^{r-m} \sum_{\lambda \vdash r} f_{\lambda,m} \langle s_{\lambda}\rangle_{\beta},
$$
hence the desired Schur function expansion (\ref{F(t,a)}) follows after changing the order of summations.

\section{Asymptotic constants and correction factors}\label{App2}

\textbf{Conductance, left edge:}
To calculate $L_g=\langle\delta(1-\sum_jT_j)\rangle$, we use the following result due to Mehta \cite{Mehta2} (see p. 361):
\begin{eqnarray}\label{b1}
  \int_0^\infty\!\!\!\cdots\!\!\int_0^\infty\!\! \prod_{j=1}^{n} dx_j x_j^{\alpha -1} |\Delta(x)|^\beta
  \frac{\Theta\bigl(1 - \sum_i x_i \bigr) }{ \bigl(1 - \sum_i x_i \bigr)^{1-\gamma}}
  = \frac{\Gamma(\gamma)}{\Gamma[\gamma + \alpha n + \frac{\beta}{2} n (n-1)]} \prod_{i=1}^n \frac{\Gamma[\alpha + \frac{\beta}{2} (n-i)]\, \Gamma(1 + \frac{\beta}{2} i) }{
  \Gamma( 1 + \frac{\beta}{2}) }.
\end{eqnarray}
Making then use of the known identity $\lim_{\gamma\rightarrow0^{+}} \gamma(1-x)^{\gamma-1} \Theta(1-x) \Theta(x) = \delta(1-x)$
and noting that $\gamma\Gamma(\gamma)= \Gamma(1+\gamma)\rightarrow1$ at $\gamma\rightarrow0$, one can readily get from (\ref{b1}) the following result (valid at any $\beta>0$):
\begin{eqnarray}\label{b2}
  \int_0^1\!\!\!\cdots\!\!\int_0^1\!\! d[T] \prod_{j=1}^{n}  T_j^{\alpha -1} |\Delta(T)|^\beta
  \delta\bigl( 1 - \sum_i T_i \bigr) = \frac{1}{\Gamma[\alpha n + \frac{\beta}{2} n (n-1)]} \prod_{j=0}^{n-1} \frac{\Gamma(\alpha + \frac{\beta}{2}j)\, \Gamma[1 + \frac{\beta}{2}(j+1)] }{
  \Gamma( 1 + \frac{\beta}{2}) }.
\end{eqnarray}
After dividing (\ref{b2}) with the normalization constant (\ref{Nbeta}), this yields Eq.~(\ref{L_g}) of the main text.

\textbf{Conductance, right edge:}
For considering  the limit $g\rightarrow n$ from below, we make a transformation $T_i \rightarrow 1{-}T_i$ in the integral for
$
 P_g^{(\beta)}(g) =  \mathcal{N}_\beta^{-1} \!\int\! d[T]\prod_j (1-T_j)^{\alpha-1} |\Delta(T)|^\beta \delta\bigl(n - g - \sum_i T_i\bigr)
$
and then consider $T_i$ small. Now one sees that in the region $n > g > n-1$ the upper bound of the integral can be replaced by $n-g$. Further rescaling $T_i\rightarrow(n-g)T_i$ yields
\begin{equation}
  P_g^{(\beta)}(g) \simeq R_g (n-g)^{(n-1)(1+\frac{\beta}{2}n)} e^{-(\alpha-1)(n-g)},
\end{equation}
where $R_g$ is then given by Eq.~(\ref{R_g}) (use result (\ref{b2}) at $\alpha=1$ and divide it by $\mathcal{N}_\beta$). The correction factor $e^{-(\alpha-1)(n-g)}$ comes from expanding $\prod_i(1-(n{-}g)T_i)^{\alpha-1}$ linearly in the exponent. Thus we have systematically expanded the log of the positive quantity $P_g^{(\beta)}(g)$ for small $(n - g)$ including the term of order $(n-g)$. Corrections are of relative order $(n-g)^2$.

\textbf{Conductance, $1 < g < 2$:}
Since we know $P_g^{(\beta)}(g)$ in the interval $0< g < 1$ exactly, one can consider asymptotics in the interval $1 < g < 2$ for $g\rightarrow 1 $. To this end, we first arrange the integration variables as $T_1<T_2<\cdots<T_n$, then rescale $T_k \rightarrow T_n T_k$ for $k=1,2,\ldots,n{-}1$, and finally perform the $T_n$ integration. As a result, one arrives at the following exact representation of the conductance distribution at $0<g<2$:
\begin{equation}\label{pg2int}
  P_g^{(\beta)}(g) = g^{\alpha n+\frac{\beta}{2}(n-1)n-1}[L_g - \widetilde{P}(g)],
\end{equation}
where
$
  \widetilde{P}(g) = (n/\mathcal{N}_\beta) \int d[T] |\Delta(T)|^\beta \prod_{i=1}^{n-1} T_i^{\alpha-1}(1-T_i)^\beta  \bigl( 1+\sum_{i=1}^{n-1} T_i \bigr)^{-n\alpha - n(n-1)\beta/2} \Theta\bigl(g-1 - \sum_{i=1}^{n-1} T_i\bigr).
$
In the particular case of $n=2$, this integral is one-dimensional and can be easily evaluated explicitly, resulting in (\ref{pg2}).  At any $n>2$, $\widetilde{P}(g)$ can be handled with the same methods as before, yielding the following asymptotic behavior at small $g-1$:
\begin{equation}
  \widetilde{P}(g) \simeq \Theta(g-1) H (g-1)^{\alpha(n-1)+\frac{\beta}{2}(n-2)(n-1)} e^{-J(g-1)},
\end{equation}
where the constants $H$ and $J$ are given by
\begin{equation}
  H = \frac{n[\alpha + \frac{\beta}{2}(n-1)]}{\Gamma[1+\alpha(n-1) + \frac{\beta}{2}(n-2)(n-1)]}
  \prod_{j=1}^{n-1} \frac{\Gamma[1+ \alpha + \frac{\beta}{2}(n+j-1)]}{\Gamma[1+\frac{\beta}{2}(j+1)]}
\end{equation}
and
$J = \frac{(n-1)(\alpha+(n-2)\beta/2)(\alpha n + \beta + (n-1)n\beta/2)}{
 1+\alpha(n-1) + (n-2)(n-1)\beta/2}$. These are the expressions used to make plots on Fig. 4.

\textbf{Shot-noise, right edge:}
Shifting all $T_i \rightarrow \frac{1}{2}+T_i$ gives
$
 P_p^{(\beta)} (p) = \mathcal{N}_\beta^{-1} \!\int\! d[T] \prod_j (\frac{1}{2}{+}T_j)^{\alpha-1} |\Delta(T)|^\beta \delta\bigl( \frac{n}{4} - p - \sum_i T_i^2\bigr),
$
where the integration is now over the $n$-dimensional cube centered at origin: $-\frac{1}{2}<T_i<\frac{1}{2}$. Considering the right edge of the support of $P_p^{(\beta)}(p)$, thus $\frac{n}{4}-p\to0$, one can then rescale $T_i \rightarrow T_i \sqrt{\frac{n}{4}-p} $ and obtain:
\begin{equation}
  P_p^{(\beta)} (p) \simeq R_p \bigl(n/4-p\bigr)^{n/2 + n(n-1)\beta/4 - 1} e^{- (n/4-p) Y}
\end{equation}
where
$
 R_p = (2^{(1-\alpha)n}/\mathcal{N}_\beta) \int_{-\infty}^{\infty}\!\cdots\!\int_{-\infty}^{\infty} d[T] |\Delta(T)|^\beta \delta (1 - \sum_i T_i^2).
$
This integral can be further reduced to a Gaussian type integral found in Mehta \cite{Mehta2} (see p. 354). After some algebra, we finally arrive at
\begin{equation}
  R_p = \frac{ \pi^{n/2} }{ 2^{ (\alpha-1)n + \beta(n-1)n/4 } \Gamma(r_p+1) }
  \prod_{j=0}^{n-1} \frac{\Gamma[1 + \alpha + \frac{\beta}{2}(n+j-1)] }{
  \Gamma(1+\frac{\beta}{2}j)\ \Gamma(\alpha+\frac{\beta}{2}j)} ,
\end{equation}
where $r_p$ is given by (\ref{lr_p}). The correction factor $e^{-(n/4-p)Y}$, with $Y = 2(\alpha-1) - 2(\alpha-1)^2/[1+\frac{\beta}{2}(n-1)]$, is found as before by expanding the rest in powers of $\frac{n}{4} -p$ and retaining the linear part in the exponent.

\textbf{Shot-noise, left edge:}
For $p \rightarrow 0$, one gets positive contributions to $P_p^{(\beta)}(p)$ from all $2^n$ corners of the integration cube. As function $T(1-T)$ is not monotonous, it is more convenient to treat the contributions from all the corners separately. To this end, we make for $(n-m)$ variables $T_k$ the transformation $T_k \rightarrow  (1-T_k)$ and then let all $T_i$ run from 0 to $\frac{1}{2}$. This gives
\begin{eqnarray}
 \nonumber
 P_p^{(\beta)} (p) &=& \frac{1}{\mathcal{N}_\beta} \sum_{m=0}^n {{n}\choose{m}} \int_0^{1/2}
 d[T] \prod_{i=1}^{m} \prod_{k=m+1}^{n} T_i^{\alpha-1} (1-T_k)^{\alpha-1} |1-T_i-T_k|^\beta
 \\ \nonumber
 && \times \prod_{1\leq i<j\leq m} |T_i-T_j|^\beta \prod_{m+1\leq k<l\leq n} |T_k-T_l|^\beta \delta \bigl( p - \sum_i T_i (1 - T_i)\bigr).
\end{eqnarray}
The factor ${{n}\choose{m}}$ appears since one has ${{n}\choose{m}}$ equivalent corners, and $2^n=\sum_{m=0}^{n}{{n}\choose{m}}$. Scaling $T_i \rightarrow p T_i$ yields
\begin{equation}
 P_p^{(\beta)} (p)  \simeq \sum_{m=0}^n  {{n}\choose{m}} Q(m)  p^{\gamma(m)-1} {\rm e}^{p E(m)},
\end{equation}
with $\gamma(m) = n + m (\alpha-1) + \frac{\beta}{2}[m(m-1) + (n-m) (n-m-1)] $ and
\begin{eqnarray}
 Q(m) &=& \frac{1}{\Gamma[\gamma(m)]} \prod_{j=0}^{n-1}
 \frac{ \Gamma[1+\alpha+\frac{\beta}{2}(n+j-1)] }{ \Gamma(\alpha+j \frac{\beta}{2}) }
 \prod_{j=0}^{m-1} \frac{\Gamma[\alpha + j \frac{\beta}{2}] \, \Gamma[1+\frac{\beta}{2}(j+1)] }{ \Gamma[1+\frac{\beta}{2}(n-m+j)] \, \Gamma[1+\frac{\beta}{2}(n-m+j+1)]},
 \\
 E(m) &=& 3 - \alpha + \beta (n-2m-1) + \frac{2m}{\gamma(m)}
 [\alpha-1+\beta(2m-n)] [\alpha+\frac{\beta}{2} (m-1)].
\end{eqnarray}
For the given even $n=2k$ or odd $n=2k+1$, $k\ge1$, the function $\gamma(m)$ has a minimum at $m=k$ that gives the leading exponent $\ell_p = \gamma(k)-1$ stated in Eq.~(\ref{lr_p}), and $L_p={n\choose{k}}Q(k)$.

\end{widetext}


\begin{thebibliography}{30}
\bibitem{clambda}
 In the mathematical literature \cite{MacDonald} the inverse of $c_\lambda$ is known as the product of all hook lengths of the partition $\lambda$.
\bibitem{b_j}
 Our analysis suggests that for those partitions which have all the $a_j$'s distinct, the corresponding $b_j$'s are given by $b_j=a_j-\lambda_j(2/\beta-1)$. In the general case, finding similar relation between these two objects is an interesting open problem.
\bibitem{pfaff}
 Pfaffian, or quaternion determinant, of an antisymmetric matrix $A$ is defined as follows:\cite{Mehta2} $\mathrm{Pfaff\,}(A)=\sqrt{\det(A)}$. It is important to note that a Pfaffian is analytic in the matrix entries. Also, in the particular case of unitary symmetry, $\beta=2$, one generally has $A=i\sigma_2\otimes\tilde{A}$, with $\sigma_2$ being the corresponding Pauli matrix in the quaternion subspace, thus $\mathrm{Pfaff\,}(A)\equiv\det(\tilde{A})$ in this case.
\end{thebibliography}

\end{document}